\documentclass[twocolumn,showpacs,prb]{revtex4}
\usepackage{bm}
\usepackage{epsfig}

\begin{document}

\title{Resonant spin-dependent electron coupling\\ in a
III-V/II-VI heterovalent double quantum well}

\author{A.~A.~Toropov} %\email{toropov@beam.ioffe.rssi.ru}
\author{I.~V.~Sedova }
\author{S.~V.~Sorokin}
\author{Ya.~V.~Terent'ev}
\author{E.~L.~Ivchenko}
\author{S.~V.~Ivanov}
\affiliation{Ioffe Physico-Technical Institute, Russian Academy of
Sciences, St. Petersburg 194021, Russia}

\begin{abstract}
We report on design, fabrication, and magnetooptical studies of a
III-V/II-VI hybrid structure containing a
GaAs/AlGaAs/ZnSe/ZnCdMnSe double quantum well (QW). The structure
design allows one to tune the QW levels into the resonance, thus
facilitating penetration of the electron wave function from the
diluted magnetic semiconductor ZnCdMnSe QW into the nonmagnetic
GaAs QW and vice versa. Magneto-photoluminescence studies
demonstrate level anticrossing and strong intermixing resulting in
a drastic renormalization of the electron effective $g$ factor, in
perfect agreement with the energy level calculations.
\end{abstract}

%75.70.Cn Magnetic properties of interfaces multilayers, superlattices, heterostructures
%78.67.Pt Optical properties of low dimensional, mesoscopic, and nanoscale materials and structures
%Multilayers; superlattices
%75.50.Pp Magnetic semiconductors
%85.75.Mm Spin polarized resonant tunnel junctions

\maketitle

The great majority of currently used semiconductor device
heterostructures are {\it isovalent}, i.e., they involve compounds
of the same chemical group. The design and fabrication of {\it
heterovalent} heterostructures, including compounds of different
groups, are hampered because of the lack of precise data on the
properties and technology of heterovalent interfaces. Particularly
discouraging are the presence of polarization charges at the
interface and poor technological reproducibility of such basic
interface properties as band offsets etc. On the other hand,
certain useful characteristics of the heterovalent structures are
unachievable in the isovalent ones. One known example is the
reduced holes leakage in mid-infrared optoelectronic devices based
on InAs, due to the realization of a huge valence band offset at
an InAs/CdSe interface.\cite{Ivanov}

Another opportunity can be a flexible engineering of
magneto-electronic and magneto-optical properties in a III-V/II-VI
hybrid structure involving a high-quality III-V part, e.g., a
GaAs/AlGaAs quantum well (QW), and a diluted magnetic
semiconductor (DMS) II-VI part, e.g. a ZnCdMnSe/ZnSe QW. Such
structures can combine large solubility of magnetic ions in the
II-VI DMS \cite{Furdyna} and high electron mobilities as well as
long electron spin relaxation times in a non-magnetic III-V
part.\cite{Kikkawa} This combination can be useful for both
fundamental studies of spin-polarized two-dimensional electron gas
and device applications in the rapidly growing field of
spintronics.

In this paper we report on the realization of a double QW
structure, where a GaAs/AlGaAs QW is electronically coupled with a
DMS ZnCdMnSe/ZnSe QW through a heterovalent AlGaAs/ZnSe interface.
We show that the proper structure design allows one to achieve
resonant tunnelling conditions, which facilitates extension of the
electron wave function in the II-VI DMS region, resulting in giant
values of the effective $g$ factor in both the GaAs-like and
ZnCdMnSe-like electronic states.

\begin{figure} [b]
  \centering{\epsfbox{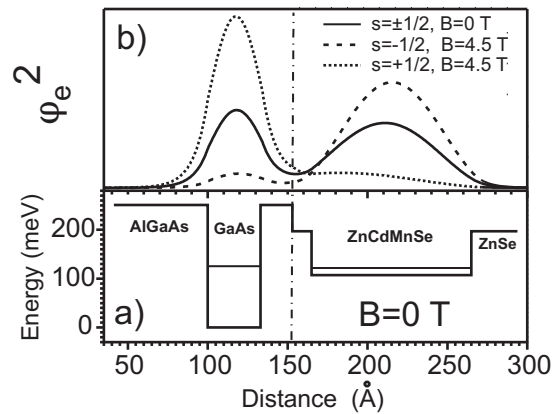}}
  \caption{ \label{f1} (a) Conduction band line-ups of the double
  QW sample. Geometrical parameters correspond to the experimental
  sample with the 3.4 nm wide GaAs QW. (b) Spin-up (dotted curve) and spin-down
  (dashed curve) squared electron wave functions calculated for the
  magnetic field 4.5 T. Solid curve shows the zero-field wave
  function.
  }
\end{figure}

The sample design and principles of operation are illustrated in
Fig.~1. Figure~1a shows schematically the conduction band line-ups
of the structure. The QW parameters are chosen in such a way that
at zero magnetic field the lowest confined electron level in the
ZnCdMnSe QW is nearly resonant with the lowest electron level in
the GaAs QW. The calculated squared envelope wave function of the
lowest-energy electron state in the coupled QWs at zero magnetic
field is plotted in Fig.~1b (solid curve). The electron
probability is almost equally distributed between the two QWs. The
primary effect of a relatively low external magnetic field applied
in the Faraday geometry is a giant Zeeman splitting in the DMS QW,
caused by the exchange interaction between electrons and Mn$^{2+}$
ions.\cite{Furdyna} As a result, the magnetic field removes spin
degeneracy, pushing the electron level with spin component
$s=-1/2$ down and the level $s= +1/2$ up. Due to the splitting,
the interwell coupling strength is different for the electrons
with different spin orientations. The spin asymmetry is well seen
in Fig.~1b showing the ground state electron wave functions for
$s= \pm 1/2$ at the magnetic field $B=4.5$~T. The calculation is
performed by using the envelope function approach as well as the
mean-field approximation while describing Brillouin-like
paramagnetic behavior of the Mn$^{2+}$ ions.\cite{Furdyna}

\begin{figure} [t]
  \centering{\epsfbox{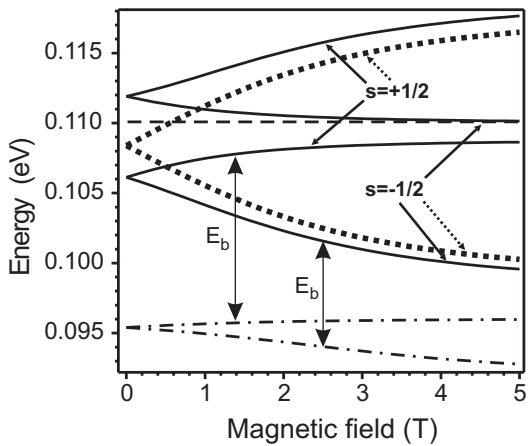}}
  \caption{ \label{f2} Electron level variation with the magnetic field
  in the double QW structure. The structure parameters are the same as in Fig.~1.
  Dashed and dotted curves represent the levels in isolated
  GaAs and ZnCdMnSe QWs, respectively. Solid curves
  show the four levels in the coupled QWs, while a pair of dash-and-dotted curves show the
  lowest spin-up and spin-down electron levels with subtracted
  exciton binding energies.
  }
\end{figure}

The calculated electron levels versus magnetic field are shown in
Fig.~2. The diamagnetic shift of the levels as well as the Zeeman
level splitting due to the intrinsic $g$ factors are neglected
since they are weak as compared with the effect of exchange
interaction with the magnetic ions. The dashed and dotted curves
in Fig.~2 illustrate the variation of energy levels in the
corresponding uncoupled single QWs. Within the used approximation
the level in the isolated GaAs QW remains spin degenerate (dashed
curve). At zero magnetic field, due to the interwell coupling, the
levels are repulsed, their splitting increases by a factor of 3,
but the double degeneracy is not lifted. At low magnetic field,
the spin splitting of each level is linear in $B$ and, due to the
strong level mixing, the splittings are comparable. As the field
increases, the single-QW levels with $s=1/2$ first approach each
other, merge at $B \approx 0.6$ T and then move apart. This
explains a drastic anticrossing of these levels when the interwell
coupling is switched on. In contrast, the increasing magnetic
field loosens the interwell coupling of the levels with $s=-1/2$.
This resonant magnetic-field-induced control of level mixing
should manifest itself in magneto-optical spectra not only in a
giant splitting between the lowest levels with $s=\pm 1/2$ but
also in a remarkable red-shift of their center-of-mass.
Penetration of the heavy hole states into the ZnCdMnSe QW is
prohibited due to the huge valence band offset at the GaAs/ZnSe
interface ($\sim$ 1.1 eV) as well as the larger value of the
effective mass.

Realization of the proposed design relies on the controlled
fabrication of a high-quality interface between the III-V and
II-VI parts. The (Al)GaAs/ZnSe interface is at present most
studied among other heterovalent interfaces. Its technology was
thoroughly developed for the growth of ZnSe-based optoelectronic
devices on GaAs substrates.\cite{Kato} More recently, injection of
spin-polarized electrons through a GaAs/ZnSe heterointerface has
been realized in an (In)GaAs/AlGaAs QW light-emitting diode with a
II-VI DMS spin aligner grown on top.\cite{Fiederling,Jonker}
Furthermore, photoluminescence was detected from an AlAs/GaAs/ZnSe
QW with a heterovalent interface.\cite{Kudelski} However, to the
best of our knowledge, the electron resonant tunnelling through a
heterovalent interface has not been observed so far.

The samples were grown by molecular-beam epitaxy on GaAs(001),
with the III-V and II-VI growth chambers being connected via an
ultra-high vacuum transfer module. The GaAs QW sandwiched between
Al$_{0.3}$Ga$_{0.7}$As barriers was grown at a substrate
temperature T$_S$=580--600$^{\circ}$C and an As/Ga flux ratio as
low as possible. The top barrier was as thin as 2 nm. It was
capped by one monolayer of GaAs to prevent contamination of the
AlGaAs surface in the transfer chamber. The grown AlGaAs/GaAs QW
structure was cooled down with the (2$\times$4)As reconstruction.
Thereafter the structure was transferred to the II-VI chamber
where it was heated up to 280$^{\circ}$C keeping the
(2$\times$4)As reconstruction unchanged. The II-VI growth was
initiated under the surface exposure to Se flux, which immediately
changed the surface reconstruction to (1$\times$1). The ZnSe
growth occurred under the (2$\times$1)Se-stabilized surface
conditions. The II-VI part contained a 10-nm-thick
Zn$_{0.85}$Cd$_{0.10}$Mn$_{0.05}$Se QW embedded between 1.2- and
20-nm-thick ZnSe barriers on bottom and top, respectively.

The thickness of the combined AlGaAs/ZnSe barrier between the GaAs
and ZnCdMnSe QWs totals 3.2 nm. The AlGaAs layer is inserted in
order to move the heterovalent interface with presumably enhanced
density of defects aback from the GaAs QW. The ZnSe spacer is
needed for proper II-VI growth initiation and for preventing Mn
diffusion in the III-V part, since even low content of Mn in III-V
compounds damages their optical quality. The total barrier
thickness governs the transfer integral between the single-QW
electron wave functions and, hence, the strength of the interwell
coupling.

A saturation value of the Zeeman splitting for electrons confined
within the ZnCdMnSe QW lies in the range of 15$\div$20~meV. This
means that the zero-field interlevel energy spacing should be
preset to within 10 meV. The fulfillment of this requirement is
complicated by the drastic dependence of the conduction band
offset (CBO) at a GaAs/ZnSe interface on growth conditions. When
the interface growth regime changes from Zn-rich to Se-rich, CBO
has been found to vary from 100 meV to 600 meV.\cite{Nicolini} We
fixed both the interface and the ZnCdMnSe QW growth conditions and
grew a set of structures with different widths of the GaAs QW,
which was controlled by growth rate calibrations and transmission
electron microscopy measurements. Most intriguing results have
been obtained on the structure with a 3.4-nm-thick QW.

\begin{figure} [t]
  \centering{\epsfbox{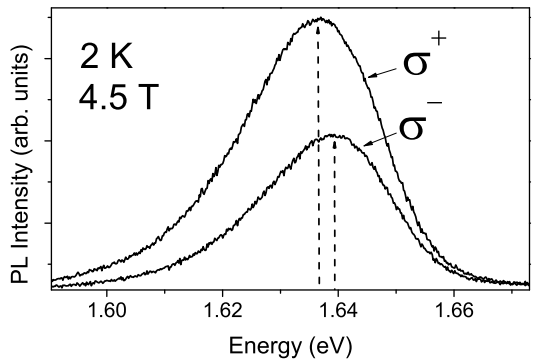}}
  \caption{ \label{f3} $\sigma^+$ and $\sigma^-$ polarized
   PL spectra measured at 4.5 T in the sample with a 3.4 nm wide
   GaAs QW.
  }
\end{figure}

\begin{figure} [b]
  \centering{\epsfbox{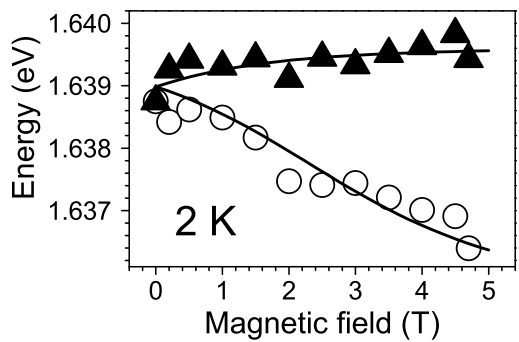}}
  \caption{ \label{f4} Energy of the PL peaks corresponding to $|-1/2, 3/2 \rangle$
  (open circles) and $|1/2, -3/2 \rangle$ (solid triangles) excitons. Solid
  lines represent a theoretical fit (see text).
  }
\end{figure}

To study the effects of interwell electronic coupling we measured
low-temperature spectra of GaAs QW excitonic photoluminescence
(PL) in a magnetic field applied in the Faraday geometry. A 514 nm
line of an Ar$^+$ laser was used as an excitation source. Figure 3
shows the PL spectra of circularly polarized emission components
measured in the sample with a 3.4 nm wide GaAs QW at 4.5 T.
According to the optical selection rules, the emission components
$\sigma^{\pm}$ are due to the radiative recombination of the
dipole-active excitons $|-1/2,3/2\rangle$ and $|1/2,-3/2\rangle$,
respectively. Here we use the notation $|s, m \rangle$ for an
exciton with the electron spin $s = \pm 1/2$ and the hole angular
momentum component $m = \pm 3/2$. The PL spectral peaks are split
by $\sim$3 meV with the lowest-energy peak being $\sigma^+$
polarized. The peak energy positions are plotted in Fig.~4 as a
function of the magnetic field. These dependencies reflect neither
the Zeeman splitting expected for a conventional GaAs/AlGaAs QW
nor the symmetrical giant splitting typical for a single DMS QW.
Indeed, for a single 3.4 nm GaAs-based QW, the expected spin
splitting at 4.5~T would be as small as 0.2$\div$0.3 meV, taking
into account the values $g_e \sim 0.1$ and $g_{hh} \sim - 1.6$ for
the electron and heavy-hole $g$ factors.\cite{Snelling} On the
other hand, the spin splitting between the $|-1/2,3/2>$ and
$|1/2,-3/2\rangle$ exciton states is quite asymmetric, namely, the
exciton level $|1/2,-3/2\rangle$ is rather stable so that the main
part of the magnetic-field-induced splitting is contributed by a
red shift of the $|1/2,-3/2\rangle$ exciton level. Therefore we
can unambiguously attribute the character of the observed PL band
splitting to the effect of resonant coupling between electronic
states in the nonmagnetic and DMS QWs. This interpretation is
confirmed by the fact that no remarkable splitting has been
observed in the off-resonant samples with thicker GaAs QWs, e.g.,
with a 6-nm-thick QW, where the electronic levels in the
nonmagnetic and magnetic QWs are remote far enough.

\begin{figure} [t]
  \centering{\epsfbox{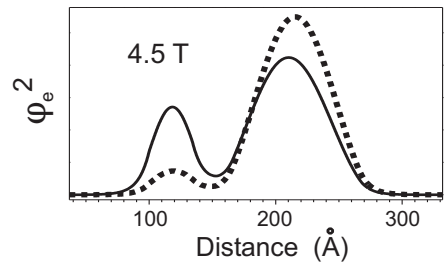}}
  \caption{ \label{f5} Electron density $\varphi^2_e(z)$
  for the spin-down lowest-energy electron state calculated at 4.5 T
  either neglecting (dotted curve) or taking into account
  (solid curve) the Coulomb-attraction-induced redistribution of the
  electron envelope function.
   }
\end{figure}

To describe the experimental data quantitatively, we have
calculated the spin-dependent energies of $|1/2,-3/2\rangle$ and
$|-1/2,3/2\rangle$ excitons as a function of magnetic field in the
coupled QW system. The band gap of the ZnCdMnSe quaternary solid
alloy was interpolated using the known band gap dependence for
Zn$_{1-x}$Cd$_{x}$Se\cite{Lozykowsky} and
Zn$_{1-y}$Mn$_{y}$Se.\cite{Bylsma} As regards the ZnCdMnSe/ZnSe
interface, we assumed that 75\% of the total band offset falls on
the conduction band. CBO at the GaAs/ZnSe interface was considered
as the only fitting parameter. For the $s$-$d$ exchange integral
and the effective concentration of Mn$^+$ spins, we took the
values $N_0 \alpha=0.26$, as in pure ZnMnSe,\cite{Twardowski} and
0.03, respectively. The dielectric constant $\epsilon = 11$ was
used as an average between those of GaAs ($\epsilon \approx 13$)
and ZnSe ($\epsilon \approx 9$).

While calculating the exciton energies we used the self-consistent
variational method and chose factorized exciton envelope functions
similar to the procedure applied in Ref.~\cite{Ivchenko}. The
probe exciton envelope function was taken in the form
\begin{equation} \label{psif}
\Psi = \varphi_e(z_e) \varphi_h(z_h) f(\rho)\  ,
\end{equation}
where $\varphi_e(z)$ and $\varphi_h(z)$ are the single-particle
electron and hole envelope functions, the envelope function
$f(\rho)$ describes the in-plane electron-hole relative motion,
$z$ is directed along the growth direction, and $\rho$ denotes the
electron-hole in-plane distance. The hole envelope $\varphi_h$ is
fixed due to strong confinement in the GaAs QW, its dependence on
the magnetic field can be ignored. The electron envelope function
is more flexible, and it can be redistributed, as compared with a
single-electron state, between the two coupled QWs due to the
electron-hole Coulomb attraction. The self-consistent solution of
the coupled Shr\"{o}dinger equations for the envelopes
$\varphi_e(z)$ and $f(\rho)$ was found numerically. The relevance
of this approach is illustrated in Fig.~5 showing the shape of
$\varphi^2_e$ at 4.5 T, calculated either self-consistently or
neglecting the Coulomb-attraction-induced redistribution of the
electron probability. The Coulomb attraction results in a
remarkable increase of probability to find the electron in the
GaAs QW. As a consequence, the self-consistent exciton binding
energy increases by about 30\%.

To compare the theory with the PL experimental data, the Stokes
shift adopted as $0.6 \Delta_{\rm PL}$, $\Delta_{\rm PL}$ being
the full width at half maximum of the PL peak,\cite{Donnel} was
subtracted from the calculated free-exciton energy. The best
theoretical fit obtained in that way is shown in Fig.~4 by solid
lines. To illustrate the effect of electron-hole Coulomb
interaction on the exciton spin splitting we depicted in Fig.~2,
by a pair of dash-and-dotted curves, the spin-up and spin-down
lowest-energy electron levels reduced by the self-consistent
exciton binding energy. The difference between the curves gives
the actual exciton splitting. In agreement with the above analysis
the spin-up exciton level is only weakly dependent on the magnetic
field. The self-consistent spin-down level approaches, at high
magnetic fields, the indirect exciton formed by an electron
confined within the ZnCdMnSe QW and a hole confined within the
GaAs QW. It is important to stress that, in comparison with the
direct (intrawell) exciton, the indirect exciton is characterized
by the weaker electron-hole Coulomb interaction and smaller
binding energy. The latter effect tends to reduce the exciton spin
splitting as compared with the spin splitting of the
single-electron levels. In particular, at 4.5 T the exciton spin
splitting of 3 meV corresponds to the single-electron spin
splitting of 8.8 meV. The difference would be even larger if the
calculation did not take into account the Coulomb induced
redistribution of electrons and the pronounced difference between
the electron effective masses in GaAs (0.067 m$_0$) and ZnCdMnSe
($\sim$ 0.16 m$_0$).

The best fit is achieved, assuming the GaAs/ZnSe CBO be equal to
185 meV. This value corresponds to a "mixed" interface, neither
Zn- nor Se-rich, in agreement with the short surface exposure to
Se at the initial stage of the II-VI growth. This procedure was
performed intentionally to bring the QW levels into a resonance in
a structure with suitable QW widths. The mixed nature of the
interface can also be beneficial for partial elimination of dipole
charges at the heterovalent interface, due to averaging the
contributions from Zn- and Se-rich microscopic regions having
opposite dipole polarities.

In conclusion, we have demonstrated resonant electronic coupling
through a heterovalent AlGaAs/ZnSe interface in an optical-quality
double QW with the DMS II-VI part. The structure design allows one
to resonantly enhance penetration of the nonmagnetic QW electron
wave function into the DMS region and enhance the QW electron $g$
factor by more than one order of magnitude. Such structures are
especially beneficial for exciton optical studies, since the
electron wave function at resonance has a minimum (see the dashed
curve in Fig.~1b) at the heterovalent interface with a presumably
large density of defects which otherwise could mess up the
excitonic properties. Another potential advantage of these hybrid
structures is a possibility to insert a similar double QW in a
$p$-$i$-$n$ or Schottky diode, allowing thus an electric control
of the electron spin polarization.

\section*{Acknowledgments}
    This work was supported by the Program of the Ministry of Sciences of RF "Physics of Solid
    State Nanostructures", INTAS (01-2375), VW Foundation and RFBR (02-02-17643,
    03-02-17566). S.V.I. is grateful to RSSF.

\end{document}